\documentclass[a4paper,12pt]{article}

\newtheorem{proposition}{Proposition}[section]

\newtheorem{remark}{Remark}[section]

\usepackage{amssymb}
\title{On Boundary Control Problems in Slow Processes for Piezothermoelastic Plates}
\author{\centerline{A. Montanaro}\\
\centerline{Dip. Metodi e Modelli Matematici per le Scienze Applicate}\\
\centerline{Universit\`a di Padova}\\
\centerline{via Trieste, 63  -- 35121 Padova - ITALY}\\
\centerline{montanaro@dmsa.unipd.it}\\
\centerline{tel. +39 049 8271306  $\quad$ Fax +39 049 8271333}}

\begin{document}
\date{}
\maketitle

\begin{abstract}
We consider a piezothermoelastic panel occupied by a material of hexagonal crystal class.   We study the response when the boundary conditions vary very slowly with time
and one of the bounding faces is subject to thermal exposure.
We show that in some cases the temperature on the other bounding face can be controlled by the difference of electric potential between the faces.
\end{abstract}


\section{Introduction}
\subsection{Premise}
Piezoelectricity is the property  
of generating an electric field (mechanical stress)
 in response to an applied mechanical stress (electric field); 
pyroelectricity is the property of generating an electric field (temperature change) by a temperature change (electric field).
 
 For a piezothermoelastic body a natural problem, useful for practical applications, is to study the electromechanical effects due to a prescribed temperature on part of the boundary.
 
 The present paper is a starting study of boundary control problems 
  in plate-like bodies exhibiting pyroelectricity.  This in order to theoretically establish whether and under which boundary conditions a given boundary temperature and/or electric potential can be usefully employed to obtain some type of control, e.g. on temperature and/or electric potential, in a given material surface. This theoric study may be useful with regard to real panels  subject to sun exposure or lying in contact with an external  heat source, in order e.g. to passively exploit this boundary condition.

Here we define first a general boundary control problem and then study the particular problem of a plate occupied by a material exhibiting piezoelectric and pyroelectric properties. 
\subsection{Static Boundary Value Problem for a Piezothermoelastic body}

Here we adopt the summation convention and comma notation for partial derivatives, so that $\; a^ib_i=\sum^n_1 a^ib_i\,$ and
$\;f_{,\,i}=\partial f/\partial x_i $.

Consider a piezothermoelastic body ${\mathcal B}$ that, in the reference configuration, occupies a region $V$ with boundary surface $S$.
The deformation of the body is described by $$y_i=y_i(X_A)=\delta_{i}^{B} X_B+u_i(X_A) \, ,$$
where $y_i$ denote the spatial coordinates and $X_L$ the reference coordinates of material points with respect to the same Cartesian coordinate system.

The Piola-Kirchhoff stress tensor, electric displacement vector, and heat flux vector are respectively given by constitutive functions  
\begin{equation}  \label{eq:tLk}
t_{Ll}=\hat t_{Ll}(\theta, \, E_i,\, E_{AB}, X_A)\, , 
	\end{equation}
\begin{equation}  \label{eq:DL}
D_{L}=\hat D_{L}(\theta, \, E_i,\, E_{AB}, X_A)\,, \quad \;q_{L}=\hat q_{L}(\theta, \, E_i,\, E_{AB}, X_A)\,,
\end{equation}
	where $\;\theta, \;  \phi, \; E_{i}=-\phi_{,\,i}, \;  E_{AB}=(y_{i,A}y_{i,B}-\delta_{AB})/2\;$
	are the absolute temperature, electric potential, electric field and strain tensor, respectively.

 Balance law of linear momentum, Maxwell's equation, and balance law of conservation of energy, respectively lead to the equilibrium relations
\begin{equation}                 \label{eq:locEq}
t_{Ll, \,L}+\rho_of_l=0 \,, \qquad 	D_{L, \,L}=\rho_e \, , \qquad  	q_{L, \,L}=\rho_o \gamma \, ,
\end{equation}
 where $\,\rho_o\,$ is the mass-density in the reference configuration,  $\,f_l\,$ is the body force per unit mass,   $\,\rho_e\,$ is the body free charge density, and
 $\,\gamma\,$ is the body heat source per unit mass. 

To describe the corresponding boundary conditions, three partitions 
$\,(S_{i1} ,  S_{i2})$, $\;i=1,\,2,\,3$, of the 
boundary surface $S=\partial {\mathcal B}$ can be assigned. 
 For mechanical  boundary conditions, displacement 
 $\, \overline {\bf u}\,$  and traction 
 $\,\overline{\bf t}\,$  per unit undeformed area 
 are prescribed, respectively, on $\,S_{11}\,$ and  $\, S_{12}$;
  for electric  boundary conditions,  electric potential 
  $\,\overline{\phi}\,$  and surface-free charge  
  $\,\overline{D} \,$   per unit undeformed area 
  are prescribed, respectively,  on $\,S_{21}\,$ and  $\, S_{22}$;
  while for thermic boundary conditions,  temperature 
   $\,\overline{T}\,$
   and normal heat flux 
    $\,\overline{q}\,$ per unit undeformed area 
  are prescribed, respectively,on $\,S_{31}\,$ and  $\, S_{32}$.
  Hence, we can write
 \begin{equation}    \label{eq:yiyi}
u_i= \overline u_i \quad {\rm on}  \quad S_{11} \, ,	\qquad t_{Li}N_{L}= \overline t_i \quad {\rm on} \quad  S_{12}  \quad  \quad {\rm ('mechanical')}\, , 
\end{equation}
 \begin{equation}    \label{eq:phi}
	\phi=\overline{\phi} \quad {\rm on}  \quad S_{21} \,	,\qquad 
	D_{L}N_{L} =\overline{D} \quad {\rm on}  \quad S_{22} \,,\qquad \quad {\rm ('electric')}	\\
\end{equation}
 \begin{equation}    \label{eq:T}
	T =\overline{T} \quad {\rm on}  \quad S_{31}\,	,\qquad 
	q_{L}N_{L}=\overline{q} \quad {\rm on}  \quad S_{32} \qquad \quad {\rm ('thermic')}\,,	\\
\end{equation}
\begin{equation}    \label{eq:SyS}
S_{i1} \cup  S_{i2}=S \,,\quad
S_{i1} \cap  S_{i2}=\emptyset \, \quad (i=1,\,2,\,3) \,,
		\end{equation} 
 where $\,{\bf N}=(N_{L})\,$ is the unit exterior normal on $\,S\,$
 and $\,T=\theta-\theta_o\,$ is the incremental temperature with respect to the temperature $\,\theta_o\,$ in the reference state.
 
The boundary value problem is then stated as:
{\it to find the solution $\,(\phi, \, T, \, {\bf u})\,$ in $\,{\mathcal B}\,$ to the constitutive relations (\ref{eq:tLk}), (\ref{eq:DL}) and field equilibrium equations  (\ref{eq:locEq}) which satisfies the boundary conditions  (\ref{eq:yiyi})-(\ref{eq:T}) for given 
$\,\overline u_i, \, \overline t_i,\, \overline{\phi} , \, \overline{D} , \, \overline{T},\, \overline{q}$.}

Of course, existence and uniqueness of the solution must be separately examined, but here are assumed.
\subsection{Boundary Control Problem}               \label{subsection:BCPs}
Let 
$\,S_o \,$ be a regular surface contained in  $\,\overline {\mathcal B}:= {\mathcal B} \cup \partial{\mathcal B}$
(possibly  $\,S_o \subset \partial{\mathcal B}$), with oriented normal unit vector ${\bf N}$, let
 $\, C\,:\, S_o \rightarrow I\!\!R\,$ be a scalar smooth function,
and let $\,\chi\,$ be any one of the ten quantities $\\$
\centerline{$\; \phi \, , \quad \theta \, ,\quad   D_{L}N_{L}\, , 
\quad q_{L}N_{L}\, , \quad u_i,\quad  t_{Li}\,N_{L} \quad(i=1,\,2,\,3) \, . $}

The {\it boundary control problem for $\chi$ on $S_o$  with goal $C$},
(abbreviated to $\,BCP \, \chi|_{S_o}=C$), is the following:
 {\it For arbitrary  $\,n \,(<10)\,$ boundary conditions of the type (\ref{eq:yiyi})-(\ref{eq:T}), 
choose the remaining $\,10-n\,$  boundary conditions in  (\ref{eq:yiyi})-(\ref{eq:T}) such that  the solution 
$\;(\phi, \, T , \, {\bf u})\;$
to the corresponding boundary value problem yields $\, \chi_{|S_o}=C$.}
 
\subsection{The Boundary Control Problems solved here}

\begin{tabular}[h]{||l||c||r||r||} \hline  \hline \multicolumn{4}{||c||}{\textbf{Boundary Control Problems for  panel ${\mathcal P}$}}\\\hline 
$BCP \; I.1.3$      &  $BCP \; II.1.3$ & $BCP \;  I.3.3$& $BCP \;  II.3.3$ \\ \hline 
$x_1$ thickness d.& $x_1$ thickness d. & $x_3$ thickness d. & $x_3$ thickness d.\\
$x_3$ polariz. d.   & $x_3$ polariz. d. & $x_3$ polariz. d. & $x_3$ polariz. d. \\ \hline 
 at $x_1=h$&  at $x_1=h$ &  at $x_3=h$&  at $x_3=h$\\  
$T=\overline{T}$  & $T=\overline{T}$  & $T=\overline{T}$  & $T=\overline{T}$ \\ 
$\phi=\overline{\phi}$ & $\phi=\overline{\phi}$  & $\phi=\overline{\phi}$  & $\phi=\overline{\phi}$ \\  
${\bf t}{\bf N}=\overline{\bf t}$ &${\bf t}{\bf N}=\overline{\bf t}$& ${\bf t}{\bf N}=\overline{\bf t}$ & ${\bf t}{\bf N}=\overline{\bf t}$ \\   \hline 
 at $x_1=-h$&  at $x_1=-h$ &  at $x_3=-h$ &  at $x_3=-h$\\    
${\bf u}=\overline {\bf u}$  & ${\bf u}=\overline {\bf u}$  & ${\bf u}=\overline {\bf u}$  & ${\bf u}=\overline {\bf u}$ \\ 
${\bf D}\cdot{\bf N}=\overline{D}$ & $\phi=\overline{\phi}_2$ & ${\bf D}\cdot{\bf N}=\overline{D}$ & $\phi=\overline{\phi}_2$\\  
${\bf q}\cdot{\bf N}=\overline{q}$ & ${\bf q}\cdot{\bf N}=\overline{q}$ &${\bf q}\cdot{\bf N}=\overline{q}$& ${\bf q}\cdot{\bf N}=\overline{q}$\\   \hline
$\phi(x_1)$ controlled &                        & $\phi(x_3)$ controlled  &              \\
$T(x_1)$ controlled &  $T(x_1)$ controlled  & $T(x_3)$ controlled   &  $T(x_3)$ controlled    \\  
$-h \leq x_1 < h $ &  $-h \leq x_1 < h $   &  $-h \leq x_3 < h $   &    $-h \leq x_3 < h $    \\ \hline \hline
\end{tabular} 

In the present paper, we solve some boundary control $BCP$, for certain linear piezothermoelastic bodies, that occupy a plate $\,{\mathcal P}\,$ infinite in extent and bounded by two parallel planes. The plate has a natural equilibrium state, i.e., with no initial field, and is occupied by a heat-conducting piezoelectric material with the symmetry of the hexagonal crystal class $\,C_{6 \nu}=6mm$, so that ferroelectric ceramics are included. 
We assume that  $\,{\mathcal P}\,$  is subject to a constant temperature on the upper face that in effect may vary slowly with time.  
On the lower face, the displacement is prescribed, as when, for example,  $\,{\mathcal P}\,$  is welded to a fixed flat body.  

We study processes which are homogeneous on each plane parallel to the boundary planes, that is, they depend only on the thickness coordinate, and, moreover, vary very slowly with time.
The precise equilibrium boundary value problems studied are summarized in the table and are completely solved once their exact solutions are determined.

In these problems, we take $n=9$ and $\,S_o \,$ any fixed plane parallel to the plane boundaries of the plate. 
In $BCP$.s $I.1.3$ and $I.3.3$ either $\chi=T$ or $\chi=\phi$, i.e., either temperature or electric potential  can be controlled on $\,S_o$.

In $BCP$.s $II.1.3$ and $II.3.3$ we have $\chi=T$, i.e. 
whatever temperature is prescribed at the upper face, the temperature can be controlled on $\,S_o\,$ by the electric potential difference between the two bounding planes.
\section{Linear Piezothermoelasticity}					\label{section:Thermo-el}
\subsection{Linear constitutive equations}							   \label{subsection:ConstEq}
The linear constitutive equations are specified below in terms of the constitutive coefficients: $\; c_{klij}=\;$ elastic moduli;
		  		  $\; e_{ikl}=\; $  piezoelectric moduli;
		  	    $\; \beta_{kl}=\;$  thermal stress moduli;
		  		  $\; \kappa^E_{kl}=\;$  dielectric susceptibility;
		  		  $\; \tilde{\overline{q}}^{}_k =\; $ pyroelectric polarizability;
		  		  $\; \varepsilon^{}_{kl} =\;$ permittivity moduli;
		  		  $\; \kappa_{kl}=\;$   Fourier coefficients;
		  		  $\; \gamma =\;$  heat capacity;
		  		  $\; \eta_o =\;$ entropy at the natural state;
		  		  $\; \rho_o \;=\;$ mass-density at the natural state.   
These coefficients, each assumed to be constant, satisfy the following symmetry conditions:
\begin{equation}             \label{eq:symmetr1}
c_{klij}\,=\,c_{ijkl}\,=\,c_{lkij}\,=\,c_{klji}\,, \qquad e_{kij}\,=\,e_{kji} \,  , 
\end{equation}
\begin{equation}             \label{eq:symmetr2}
\beta_{ij}\,=\, \beta_{ji}  \, ,
\qquad	\kappa_{kl}\,=\,\kappa_{lk}\,  , \qquad \kappa^E_{kl}\,=\,\kappa^E_{lk} \, .
\end{equation}
With respect to a natural reference state, i.e., a state free from mechanical and/or electric fields, and with constant temperature $\theta_o$,  we assume the following standard constitutive equations \cite{FVPL}-\cite[p.122]{Y:NET}, \cite{C:GSPT}), respectively for the stress tensor,  electric displacement vector and heat flux vector:

\begin{equation}             \label{eq:ConstEquat1}
t_{kl}\,=  \,	c_{klij}\, u^{}_{i,\,j} \, - \, e_{ikl}\, E_{i} \, -\,	\beta_{kl}\, T \, ,
\end{equation}
\begin{equation}             \label{eq:ConstEquat2DDDD}
D_{k} \, = \, \, e_{kij}\, u^{}_{i,\,j} \, + \, \varepsilon^{}_{ki}\,E_{i}  \,+\, \tilde{\omega}^{}_k \, T \,,  \qquad  q_{k} \, = \, -\, \kappa^{}_{kl} \, T^{}_{,\,l} \,-\, \kappa^E_{kl} \, E_{l} \,,
\end{equation}
where $\;E_i \,=\, - \phi_{,\,i}\,$,
 %
$\;T=\theta - \theta_o \; $ is the incremental absolute temperature with respect to the absolute temperature $\;\theta_o\;$ in the natural reference state.

\subsection{Field equations of equilibrium}															  \label{subsection:FieldEqsp}
The linearized field equations of equilibrium, obtained by substituting the constitutive equations (\ref{eq:ConstEquat1})-(\ref{eq:ConstEquat2DDDD}) in the balance laws (\ref{eq:locEq}) taking
$\,f_k=\rho_e=\gamma=0\,$ are given by
\begin{equation}             \label{eq:balance1bbb}
	c_{klij}\, u^{}_{i,\,jk} \, +	\, e_{ijl}\, \phi^{}_{,\,ij} \,-\,	\beta_{kl}\, T^{}_{,\,k}\,=\,0  \qquad   (l=1,\,2,\,3)\, ,
\end{equation}

\begin{equation}    \label{eq:balance2bbb}
e_{kji} \, u^{}_{j,\,ik} \, -	\, \varepsilon^{}_{kj}\, \phi_{,\,jk}\, +\, \tilde{\omega}^{}_k \, T^{}_{,\,k} \,=\,  0 \, ,  \qquad
- \kappa^{}_{kj} \, T_{,\,jk} \,+\, \kappa^E_{jk} \, \phi^{}_{,\,jk}  \,=\, 0 \, .
\end{equation}

\subsection{Use of compressed notation and matrix arrays}							   \label{subsection:ConstEqcry}
As is well known, the matrix notation consists of replacing  $\,ij\,$ or  $\,kl\,$ by $\,p\,$ or $\,q$, where $\,i,\, j, \, k, \,l\,$ take the values $\,1,\,2,\,3\,$ and $\,p,\,q\,$ take the values   $\,1,\,2,\,3\,, 4, \, 5, \,6\,$ according to the following relations: 

 \begin{tabular}{||l|c|c|c|c|c|c||} \hline
 $i j$ or $kl$ & 11 & 22 & 33 & 23 or 32  & 31 or 13  & 12 or 21  \\ $p,\,q$ & 1  & 2 & 3 & 4 & 5 & 6 \\ \hline\end{tabular} 

By virtue of the above identification, 
the constitutive equations become
\begin{equation}             \label{eq:ConstEquat1bis}
t_{p}\,=  \,	c_{pq}\, S^{}_{q} \, - \, e_{ip}\, E_{i} \, -\,	\beta_{p}\, T \, ,
\end{equation}
\begin{equation}    \label{eq:ConstEquat3bis}
D_{i} \, = \, \, e_{iq}\, S^{}_{q} \, + \, \varepsilon^{}_{ik}\,E_{i}  \,+\, \tilde{\omega}^{}_i \, T \,, \qquad 
q_{i} \, = \,- \kappa^{}_{il} \, T^{}_{,\,l} \,-\, \kappa^E_{il} \, E_{l} \,,
\end{equation}
where    
$\; S_{ij}=\frac{1}{2} \left( u_{i,\,j}+ u_{j, \, i} \right)$,
 $\; S_{ij}\,=  \,S_{p} \; {\rm when}  \; i=j=p=1,\,2,\,3$,
      $\quad 2S_{ij}\,=  \,S_{p} \quad {\rm when}  \; i \neq j, \; p=4,\,5,\,6$,
      and $\; e_{ikl}\,=  \,e_{ip}$.
\section{Hexagonal materials}  					   \label{section:Hexagonal}
\subsection{Constitutive equations for ferroelectric ceramics}  					   \label{subsection:CEmonoclinic}

The polarized ferroelectric ceramics have the symmetry of a hexagonal crystal in class $\,C_{6\,\nu}=6mm$.
Choosing $\,x_3\,$ in the polarization direction, 
assuming $\;\beta_{ij} \, =\,	0 \,$ for $\; i \neq j\;$ and putting
\begin{equation}             \label{eq:betaijjinn}
\beta_{k} \, :=\,	\beta_{kk}  \quad (\,k=1,\,2,\,3  \,) \, ,  \qquad
\left[\beta_{p}  \right] \,= \,    
          \left[ \begin{array}{llllll}
         \beta_{1},  & \beta_2, &  \beta_3, &
          0,  & 0,  &  0  \\
                  \end{array} \right] \, ,
\end{equation}
the constitutive equations (\ref{eq:ConstEquat1})  and
 (\ref{eq:ConstEquat2DDDD})$_1$  become  (cf. e.g. \cite{C:GSPT}, \cite[p.58]{T:LPPV}, and \cite{FVPL})
 \begin{eqnarray}                     \label{eqnarray:tttttt}
 t_{1} & = &c_{11}\, u_{1, \,1}+ c_{12}\,  u_{2, \,2}+c_{13} \, u_{3, \,3}+ e_{31} \,  \phi_{,\,3} \nonumber
 - \beta_{1}\,T \; , \nonumber\\
 t_{2} & = &c_{12}\,  u_{1, \,1}+ c_{11}\,  u_{2, \,2}+c_{13}\,  u_{3, \,3} + e_{31} \,  \phi_{,\,3} \nonumber
   - \beta_{2}\,T \; ,  \nonumber\\
 t_{3} & = &c_{13}\left( u_{1, \,1}+  u_{2, \,2}\right)+c_{33}\,  u_{3, \,3} +  e_{33} \,  \phi_{,\,3} 
   - \beta_{3}\,T \; ,  \\
 t_{4} & = &c_{44} \left(u_{3, \,2}+u_{2, \,3}\right) + e_{15} \,  \phi_{,\,2}  \; , \nonumber\\
 t_{5} & = &c_{44} \left(u_{3, \,1}+u_{1, \,3}\right)  +  e_{15} \,  \phi_{,\,1}  \; , \nonumber\\
 t_{6} & = &c_{66} \left(u_{1, \,2}+u_{2, \,1}\right)   \; , \qquad {\rm with} \quad	c_{66}=(c_{11}-c_{12})/2 \,,
\nonumber
 \end{eqnarray}
 \begin{eqnarray} \label{eqnarray:DDD}
D_{1} & = &e_{15}\left(u_{3, \,1} + u_{1, \,3}\right)- \varepsilon_{11}\,  \phi_{,\,1} \, + \,\tilde{\omega}^{}_1 T \,  , \nonumber\\
D_{2} & = &e_{15}\left(u_{3, \,2}+u_{2, \,3} \right)- \varepsilon_{11} \,  \phi_{,\,2} + \,\tilde{\omega}^{}_2 T  \, ,  \\
D_{3} & = &e_{31}\left(u_{1, \,1} + \,u_{2, \,2}\right) + e_{33}\,u_{3, \,3} - \varepsilon_{33} \,  \phi_{,\,3} \,+ \,\tilde{\omega}^{}_3 T  \, . \nonumber 
 \end{eqnarray}
\subsection{Field equations of equilibrium}  					   \label{subsection:fieldEquilEquat}
When the constitutive relations (\ref{eqnarray:tttttt}), (\ref{eqnarray:DDD}) are substituted in the equilibrium field equations (\ref{eq:balance1bbb})-(\ref{eq:balance2bbb})
we have  
\begin{eqnarray}             \label{eqnarray:ConstEq1111}    \nonumber
 - \beta_{1} \,T_{,\,1} \,+\, 
c_{11} u^{}_{1,\,11} 
\,+\, (c_{12} + c_{66}) u^{}_{2,\,12} 
\,+\, (c_{13} + c_{44}) u^{}_{3,\,13} \, + \\ 
\,+\,  c_{66} u^{}_{1,\,22} 
\,+\,  c_{44} u^{}_{1,\,33}               
\,+\,  (e_{31}+e_{15})\phi_{, \,13}    
 \, =\,	0 \,  \qquad \qquad
\end{eqnarray}
\begin{eqnarray}             \label{eqnarray:ConstEq1222}  \nonumber
- \beta_{2} \,T_{,\,2} 
\,+\, c_{66} u^{}_{2,\,11} 
\,+\, (c_{66} + c_{12}) u^{}_{1,\,12}  
\,+\,  c_{11} u^{}_{2,\,22} 
\,+\,  (c_{13} + c_{44}) u^{}_{3,\,23}\, +  \\
\,+\,  c_{44} u^{}_{2,\,33}                
\,+\, (e_{31}+e_{15})\phi_{, \,23} 
 \, =\,	0 \,  \qquad \qquad   
\end{eqnarray}
\begin{eqnarray}             \label{eqnarray:ConstEq1333}  \nonumber
- \beta_{3} \,T_{,\,3} \,+\, 
c_{44} u^{}_{3,\,11} 
\,+\, (c_{44} + c_{13}) u^{}_{1,\,31}  
\,+\, c_{44} u^{}_{3,\,22} 
\,+\, (c_{44} + c_{13}) u^{}_{2,\,23}\\
\,+\,  c_{33} u^{}_{3,\,33}   
\,+\, e_{15}\phi_{, \,11} 
\,+\, e_{15}\phi_{, \,22}
\,+\, e_{33}\phi_{, \,33} 
  \, =\,	0 \, \qquad \qquad   
\end{eqnarray}
\begin{eqnarray}             \label{eqnarray:ConstEq1444}     \nonumber
e_{15}u^{}_{3,\,11} \,+\,
(e_{15}+e_{31})u^{}_{1,\,13} \,+\,
 e_{15}u^{}_{3,\,22} \,+\,
(e_{15}\,+\, e_{31})u^{}_{2,\,32} \,+\,\\
 e_{33}u^{}_{3,\,33} \,  
  \,+\, \tilde{\omega}^{}_l  \,T_{,\,l} 
 \, - \, \varepsilon_{11}\phi_{, \,11} \, - \, \varepsilon_{11}\phi_{, \,22} \, - \, \varepsilon_{33}\phi_{, \,33} \, =\,	0  \,  
\end{eqnarray}
\begin{equation}    \label{eq:bal355}
- \kappa^{}_{ij} \, T_{,\,ij} \,+\, \kappa^E_{ij} \,  \phi^{}_{,\,ij}  \,=\, 0 \, .
\end{equation}
Particular forms of the solution to these equations are discussed in the next section, while general expressions are derived in the Appendix.
\section{Quasi-statics}  					   \label{section:QuasiSt}

A principal application of the present theory is to a pyroelectric plate bonded to a fixed foundation, with the upper plane face exposed to sunlight.
For this the boundary conditions include $\,(i)$ the prescription of temperature on the upper bounding plane, 
and $\,(ii)$ the condition of assigned displacement on the lower bounding plane.  
Furthermore, the prescribed boundary values may be understood to be  functions of a parameter $\,\tau\,$  which depends slowly on time: 
$$ \tau=\tau(t),\qquad |\tau'(t)|\quad {\rm small}\,.$$

Hence, we refer to equations (\ref{eqnarray:ConstEq1111})-(\ref{eq:bal355}) augmented by these slowly varying boundary conditions as a {\it boundary value problem of quasi-statics}.

\subsection{Boundary Control Problem$\;I.1.3$}  					   \label{subsection:problemIB}
\subsubsection{Statement of the problem}  					   \label{subsubsection:statProbIB}

The plate $\,{\mathcal P}\,$ is bounded by the parallel planes $\,x_1= \pm h\,$ and is coated by an infinitesimally thin electrode on the plane $\,x_1=h$, so that all its mechanical effects may be ignored. We seek solutions of the form
\begin{equation}  \label{eq:formsol}
	T=T(x_1)\,, \qquad 	\phi=\phi(x_1)\,, \qquad 	u_i=u_i(x_1) \, ,  
\end{equation}
which when substituted in  (\ref{eqnarray:ConstEq1111})-(\ref{eq:bal355})  give
\begin{equation}             \label{eq:ConstEq1111x} 
 - \beta_{1} \,T_{,\,1} \,+\, 
c_{11} u^{}_{1,\,11} \,=\,0\, , \qquad  c_{66} u^{}_{2,\,11} \, =\,	0 \, ,
\end{equation}
\begin{equation}             \label{eq:ConstEq1333x}  
c_{44} u^{}_{3,\,11} 
\,+\, e_{15}\phi_{, \,11} \,
  \, =\,	0 \, , \qquad \qquad   
\end{equation}
\begin{equation}             \label{eq:ConstEq1444x} 
e_{15}u^{}_{3,\,11}  \,+\, \tilde{\omega}^{}_1  \,T_{,\,1} 
 \, - \, \varepsilon_{11}\phi_{, \,11} \, =\,	0  \, , \quad 
 - \kappa^{}_{11} \, T_{,\,11} \,+\, \kappa^E_{11} \,  \phi^{}_{,\,11}  \,=\, 0 \, . 
\end{equation}
$BCP$ $I.1.3:\;${\it To find the solution of the form (\ref{eq:formsol}) to the field equations (\ref{eq:ConstEq1111x})-(\ref{eq:ConstEq1444x}),
subject to the ten boundary conditions
\begin{eqnarray}   \label{eqnarray:10bc} \nonumber
T(h) = \overline{T}  \, , \quad  \phi(h) = \overline{\phi} \,  ,    \quad   	
t_{1}(h) =\overline{t}_1 \,,  \quad t_{6}(h) =\overline{t}_2 \,, \quad t_{5}(h) =	\overline{t}_3 \,, \quad \\ 
 u^{}_{i}(-h)  = \overline{u}_i \quad(i=1,\,2,\,3) ,\quad 
 -D_{1}(-h) = \overline{D}   \,, \qquad 
 -q_{1}(-h) = \overline{q}  \,, \quad\quad 
\end{eqnarray}
with $\;\overline{T}, \; \overline{\phi}, \;\overline{t}_1, \;\overline{t}_2, \;\overline{t}_3,  \;\overline{u}_1,\;\overline{u}_2,\;\overline{u}_3,\; \overline{D}\;$ and $\,\overline{q}\,$
 assigned real constants.} 
\subsubsection{General solution of BCP $I.1.3$} 					              \label{subsubsection:gen-solution-P-IB}
By Proposition \ref{proposition:pr1}, with
\begin{equation}       \label{eq:putB}
	c=c_{44} , \;	e=e'=e_{15} , \; 	\omega=\tilde \omega_{1} , \; \varepsilon=\varepsilon_{11} , \;
	\kappa=\kappa_{11} , \;	\kappa'=\kappa^E_{11}, \quad \beta=0 ,
\end{equation}
\begin{equation}    \label{equation:aI.1.3}
K=k/k', \; A= c\omega, \;  B= ee'+c \varepsilon, \; 	a=AK^{-1}B^{-1}\, ,
\end{equation}
we have from (\ref{eqnarray:SystemGenSol}) that
the general solution to equations (\ref{eq:ConstEq1333x})-(\ref{eq:ConstEq1444x}) is
\begin{equation}                                \label{eq:u3}
	u_3(x_1)=-c^{-1}eKT_1e^{ax_1}+U_{31} x_1+U_{32}  \,,
\end{equation}
\begin{equation}                                \label{eq:T1}
	T(x_1)=T_1e^{ax_1}+T_2  \,,
\end{equation}
\begin{equation}                                \label{eq:phi1}
	\phi(x_1)=KT_1e^{ax_1}+F_1 x_1+F_2  \,,
\end{equation}

where 
$\; T_1, \, T_2, \,F_1, \, F_2, \, U_{3 1}, \, U_{3 2}\;$
are arbitrarily chosen smooth functions of $\,\tau$.

Further,  (\ref{eq:T1}) implies $\, T_{, \,1}=T_1ae^{ax_1}$ which together with (\ref{eq:ConstEq1111x})$_1$ yields
\begin{equation}                                \label{eq:u1}
	u_1(x_1)=a^{-1}c^{-1}_{11} \beta_1 \, T_1 \, e^{ax_1}+U_{11} x_1+U_{12} \, ,
\end{equation}
while (\ref{eq:ConstEq1111x})$_2$ gives
\begin{equation}                                \label{eq:u2}
u_2(x_1)=U_{21} x_1+U_{22}  \, ,
\end{equation}
where $\, U_{\alpha \beta}, \; \alpha, \, \beta=1,\,2$, are arbitrary smooth functions of $\, \tau$.

\subsubsection{Decomposition of $\,BCP \,I.1.3$} 					              \label{subsubsection:SplittingProbB13}
 We solve $\,BCP \,I.1.3$ by decomposition into  two parts, described below. 

{\it Part $1$ of $BCP$ $I.1.3. \quad$} We first consider the boundary conditions
\begin{eqnarray}               \label{eqnarray:6bc}  \nonumber
T(h)=\overline{T}  \,, \quad  \phi(h)=\overline{\phi} \,, \quad  t_5(h)=\overline{t}_3 \,, \qquad \qquad\\
\quad  u_3(-h)=\overline{u}_3 \, , \quad   -D_{1}(-h) \,=\, \overline{D} \,, 
 \quad -q_{1}(-h) \,=\, \overline{q}  \, .  \qquad
\end{eqnarray}
Note that by Eq.s (\ref{eqnarray:DDD})$_1$, (\ref{eq:u3})--(\ref{eq:phi1})  and (\ref{equation:a}), the 5-$th$ boundary condition above becomes       
\begin{equation}                                                 \label{eq:ccoollxxbisBpre}
 -e u_{3,\,1}+ \varepsilon \phi_{,\,1}-\omega T =-\omega T_2 + \varepsilon F_1 
 -eU_{31} =\overline{D} \,,
\end{equation}
and by (\ref{eq:ConstEquat3bis})$_2$ the 6-$th$ boundary condition above respectively
\begin{equation}                                              \label{eq:ccoollxxbisBprex}
 \kappa T_{,\,1}-\kappa'\phi_{,\,1} = -\kappa' F_1 = \overline{q} \,.  
\end{equation}

Now (\ref{eq:u3})-(\ref{eq:phi1}) satisfy (\ref{eqnarray:6bc}) provided
\begin{equation}  \label{eq:ctauugsB}
T_1\, e^{ah} \,+\, T_2 \, =\,  \overline{T} \,    ,   \qquad  \,K \,T_1\, e^{ah } \,+\,F_1 \, h \,+\,F_2 \,=\,\overline{\phi} \, ,  
\end{equation}
\begin{equation}                                                  \label{eq:bc3333implysu3B}
eF_1+cU_{31} \, =\, \overline{t}_3  \, , \qquad -c^{-1}eK T_1 \, e^{-ah}-U_{31} h + U_{32} \,=\, \overline{u}_3 \, ,
\end{equation}
\begin{equation}                                                 \label{eq:ccoollxxbisB}
-\omega T_2 -e_{}U_{31}+ \varepsilon_{}F_1 =\overline{D}  \, , \qquad -\kappa'_{}F_1=\overline{q}    \, . 
\end{equation}
By solving the above system of equations in the unknowns 
$\,(T_1, \, T_2 , \, F_1, \, F_2 , \, U_{3 \, 1}, \, U_{3 \, 2})$,
we obtain
\begin{equation}          \label{equation:coeffPIB}              
 T_2 \,=\,- \frac{c\overline{D}+e\overline{t}_3}{c \omega}    \,-\, K \frac{c\varepsilon + e^2}{c \omega k} \, \overline{q} \, , \qquad  T_1 \,=\, e^{-ah} \Big( \overline{T} - T_2\Big)  \, , 
\end{equation} 
 \begin{equation}  \label{equation:coeffFF} 
  F_1 \,=\,- K k^{-1} \, \overline{q}  \, , \qquad 
 F_2 \,=\, \overline{\phi} - h F_1 - K e^{ah}T_1    \, ,      \end{equation} 
 \begin{equation}  
 U_{31} \,=\,  \frac{1}{c}(\overline{t}_3 - e F_1) \, , \qquad 
 U_{32} \,=\,   \overline{u}_3 - h U_{31}  + \frac{Ke}{ce^{ah}}T_1 \, ,   \end{equation}

so that the solution to the first part of $BCP$ $I.1.3$ becomes in particular 
\begin{equation}               	   \label{eq:TTFFBc}           
\phi(x_1)\,=\,K \Big( e^{a(x_1-h)}- 1 \Big) \Big(\, \overline{T} \,+\, \frac{c\overline{D}+e\overline{t}_3}{c \omega}    \,+\, K  \frac{c\varepsilon + e^2}{c \omega k} \, \overline{q} \,\Big) 
\,+\, \overline{\phi}  \,- \, \frac{\overline{q}}{k'} \,  \,(x_1-h) \,,
\end{equation}
\begin{equation}               	   \label{eq:TTFFBct}           
T(x_1)\,=\,\Big(e^{a(x_1-h)}-1\Big)\Big(   \,  \frac{c\overline{D}+e\overline{t}_3}{c \omega}  \,+ \, K\,\frac{c\varepsilon + e^2}{c \omega k} \, \overline{q}  \,\Big) + e^{a(x_1-h)}\overline{T}  \,. 
\end{equation}
Hence, 
\begin{equation}               	   \label{equation:Phi-h}           
\phi(-h)=K  \Big( e^{-2ah}- 1 \Big) \Big(\, \overline{T}+\frac{c\overline{D}+e\overline{t}_3}{c \omega} + K \frac{c\varepsilon + e^2}{c \omega k} \, \overline{q} \,\Big)+\overline{\phi} + \frac{2h}{k'} \, 
\overline{q}      \,  , 
\end{equation}
\begin{equation}               	   \label{equation:TTFFBct-h}           
T(-h)\,=\,\frac{e^{-2ah}-1}{c \omega}\Big[ c\overline{D}+e\overline{t}_3+Kk^{-1}(c\varepsilon + e^2) \, \overline{q}  \, \Big] + e^{-2ah} \overline{T}  \,  ,
\end{equation}
and for $\; 0=\overline{D}=\overline{t}_3=\overline{q}\;$  we have 
\begin{equation}               	              \label{equation:Phi-h00}           
\phi(-h)\,=\, \overline{\phi}+ K \Big( e^{-2ah}-1 \Big) \overline{T}        \,  , 
\qquad T(-h)\,=\, e^{-2ah} \overline{T}  \,  ,
\end{equation}
which yield electric potential and temperature in the plane $\,x_1=-h\,$ in terms of electric potential$\, \overline{\phi}\,$ and temperature $\,\overline{T}\,$ at $\,x_1=h$.

{\it Part $2$ of $BCP$ $I.1.3. \quad$} 
Next, we use (\ref{eq:TTFFBc}), (\ref{eq:TTFFBct}) along with (\ref{eq:u3}), (\ref{eq:u1}), (\ref{eq:u2}) and (\ref{eqnarray:tttttt}) to determine the solution that satisfies the four remaining equations 
(\ref{eq:ConstEq1111x})
joined to the four remaining boundary conditions  
\begin{equation}             \label{eq:bb1bx}
t_{1}(h)\,=\,	\overline{t}_1 \, , \quad t_{6}(h)\,=\,	\overline{t}_2 \, , \quad  u^{}_{1}(-h) \,=\, \overline{u}_1 \,,  \quad  u^{}_{2}(-h) \,=\, \overline{u}_2 \,.
\end{equation}
Now by (\ref{eqnarray:tttttt}) and
(\ref{eq:u1})  we have 
$\; t_1=\beta_1T_1e^{ah}+c_{11} U_{11}- \beta_1 T\;$ and 
$\; t_6=c_{66} U_{21}$;  hence by using (\ref{eq:T1})-(\ref{eq:phi1}) the boundary conditions (\ref{eq:bb1bx}) take the form
\begin{equation}  \label{eq:bc2part0}
c_{11} U_{11}\, =\,  \beta_1 \overline{T} \,+\,\overline{t}_1   \,-\, \beta_1 T_1\, e^{ah} \, , \quad
c_{66} U_{21}\, =\, \overline{t}_2 \,,
\end{equation}
\begin{equation}                                                 \label{eq:bc2partx}
   -U_{11} h + U_{12} \,=\, \overline{u}_1 \,-\,  a^{-1}c_{11}^{-1}\beta_1 T_1  e^{-ah}  \,,
   \quad -U_{21} h + U_{22} \,=\, \overline{u}_2 \, ,
\end{equation}
with $\,T_1\,$ given by (\ref{equation:coeffPIB}).
		On solving (\ref{eq:bc2part0})-(\ref{eq:bc2partx}) for $\,U_{\alpha \beta}\,$ and substituting the resulting expressions in
(\ref{eq:u1}), (\ref{eq:u2}) we are led to the complete solution of the second part of $BCP$ $I.1.3$.

\begin{remark}                     \label{remark:a1}
We point out that in order to avoid growth as $\;  h \rightarrow +\infty \;$ of the  magnitude of the gradient
$$ |  \nabla {\bf u}| + |\nabla T |+ |\nabla \phi |$$
of any solution  $\; ({\bf u}, \; T , \; \phi ) \; $ 
to $BCP$ $I.1.3$, we assume 
\begin{equation}   \label{eq:assumption1}
\qquad \qquad \qquad \qquad \qquad a:= \frac{c_{44} \tilde\omega_{1}}{K(e_{15}^2+c_{44} \varepsilon_{11})} > 0 \, .	
\end{equation}
\end{remark} 
In fact, note that by the equalities above, if  $\, a <0$, then  
$\,u_1,\, u_3,\, T,\, \phi \, \rightarrow \infty\,$ as $\, h  \rightarrow \infty$.
\subsubsection{On controllability in BVP $I.1.3$} 					              \label{subsubsection:remarksI13}
By Eq.s (\ref{eq:TTFFBc}), (\ref{eq:TTFFBct}) we can deduce the following control property. 

 \begin{remark} 
Let $\; -h \leq x_1 < h$. For each choice of $\, \overline{u}_1,\, \overline{u}_2, \, \overline{u}_3, \, \overline{t}_1,\, \overline{t}_2, \, \overline{\phi}$,  given any three quantities  in  $\, \left\{\overline{D} , \, \overline{t}_3 , \, \overline{q}, \, \overline{T} \right\} \,$, the remaining quantity can be choosen to control either  $\,T(x_1)$  or $\,\phi(x_1)$. 
\end{remark}   

\subsection{Problem $BCP$ $II.1.3$}  					   \label{subsection:problemIC}
\subsubsection{Statement of the problem}  					   \label{subsubsection:statProbIC}
Here $\,{\mathcal P}\,$, just as in $BCP$ $I.1.3$, is bounded by the parallel planes $\,x_1= \pm h\,$ each coated by an electrode, which is infinitesimally thin, so that all mechanical effects may be ignored. We seek solutions of the form (\ref{eq:formsol})
which when substituted in  (\ref{eqnarray:ConstEq1111})-(\ref{eq:bal355})  give again Eq.s  (\ref{eq:ConstEq1111x})-(\ref{eq:ConstEq1444x}).

$BCP$ $II.1.3:\,$
{\it To find the solution of the form (\ref{eq:formsol}) to the field equations (\ref{eq:ConstEq1111x})-(\ref{eq:ConstEq1444x}),
which satisfies the ten boundary conditions  
\begin{eqnarray}   \label{eqnarray:10bcII}  \nonumber
T(h) = \overline{T}  \, , \quad  \phi(h) = \overline{\phi} \,  ,    \quad   	
t_{1}(h) =\overline{t}_1 \,,  \quad t_{6}(h) =\overline{t}_2 \,, \quad t_{5}(h) =	\overline{t}_3 \,, \quad \\ 
u^{}_{i}(-h)  = \overline{u}_i \quad(i=1,\,2,\,3) ,\quad 
\phi(-h) = \overline{\phi}_2   \,, \qquad 
 -q_{1}(-h) = \overline{q}  \,, \quad\quad
\end{eqnarray}
with
$\;\overline{T}, \; \overline{\phi}, \;\overline{t}_1, \;\overline{t}_2, \;\overline{t}_3,  \;\overline{u}_1,\;\overline{u}_2,\;\overline{u}_3,\; \overline{\phi}_2\,$ and $\,\overline{q}\,$ assigned real constants.	}
\subsubsection{General solution of $BCP$ $II.1.3$} 					              \label{subsubsection:gen-solution-P-IBB}
Insertion of (\ref{eq:formsol}) into
 the equilibrium field equations (\ref{eqnarray:ConstEq1111})-(\ref{eq:bal355}) gives Eq.s 
  (\ref{eq:ConstEq1111x})--(\ref{eq:ConstEq1444x}), whose general solution is expressed, as before, by Eq.s (\ref{eq:u3})-(\ref{eq:phi1}),
where 
$$T_1, \, T_2, \,F_1, \, F_2, \, U_{\alpha 1}, \, U_{\alpha 2} \; \, (\alpha=1,\,2, \, 3)$$
are arbitrary smooth functions of $\,\tau$.
\subsubsection{Decomposition of $BCP$ $\,II.1.3$} 					              
We solve $BCP$ $II.1.3$ by separating it into two parts, described below. 

{\it Part $1$ of $BCP$ $II.1.3. \quad$} 
We first determine the arbitrary constants in the general solution (\ref{eq:u3})--(\ref{eq:phi1}) so that the boundary conditions
\begin{eqnarray}\label{eqnarray:6bcII1.3} \nonumber
	T(h)=\overline{T} \,, \quad  \phi(h)=\overline{\phi}  \,, \quad t_5(h)=\overline{t}_3 \,, \\ 
u_3(-h)=\overline{u}_3 \, , \quad  \phi(-h)=\overline{\phi}_2 \,, \quad -q_{1}(-h)=\overline{q} \,
\end{eqnarray}
are satisfied.
Note that by (\ref{eq:ConstEquat3bis})$_2$, (\ref{eq:T1}) and (\ref{eq:phi1})   
 the  last boundary condition  becomes       
\begin{equation}                                                 \label{eq:ccoollxxbisBprexC}
 \kappa^E_{11}K T_{,\,1}-\kappa^E_{11}\phi_{,\,1} =
  -\kappa K^{-1}F_1=\overline{q}   \, .
\end{equation}
  
We have
		\begin{equation}  \label{eq:ctauugsBà}
T_1\, e^{ah} \,+\, T_2 \, =\,  \overline{T} \,   ,     \qquad
                \,K \,T_1\, e^{ah } \,+\,F_1 \, h \,+\,F_2 \,=\,\overline{\phi} \, ,
\end{equation}
\begin{equation}                                                  \label{eq:bc3333implysu3Bà}
eF_1+cU_{31} \, =\, \overline{t}_3  \, , \qquad 
-c^{-1}eK T_1 \, e^{-ah}-U_{31} h + U_{32} \,=\, \overline{u}_3 \, ,
\end{equation}  
\begin{equation}                                                 \label{eq:ccoollxxbisBà}
\,K \,T_1\, e^{-ah } \,-\,F_1 \, h \,+\,F_2 \,=\,\overline{\phi}_2 \, , \qquad
-\kappa_{}K^{-1}F_1=\overline{q}     \, ,     
\end{equation}
and by solving the above system of equations for the unknowns 
$\,(T_1, \, T_2 , \, F_1, \, F_2 , \, U_{3 \, 1}, \, U_{3 \, 2})\,$
we find expressions for the latter in terms of the boundary data.  In particular, we have
\begin{equation}                                            \label{eq:C-1-3-coeff-solT1T2}
T_1=\frac{2h}{k(e^{ah}-e^{-ah})}\Big[\overline{q} +\frac{k'}{2h}  \Big(\overline{\phi}-\overline{\phi}_2 \Big)  \Big] , \qquad   T_2=\overline{T} - e^{ah} T_1   \, , 
   \end{equation}
 \begin{equation}                                                  \label{eq:C-1-3-coeff-sol3}
F_1 \,=\, -\frac{\overline{q}}{k'} \, , \qquad F_2 \,=\,\overline{\phi} \,+\, \frac{hK}{k} \overline{q} - K e^{ah} T_1  \, . \end{equation}
Hence, by (\ref{eq:T1})--(\ref{eq:phi1}), the expressions of $\,T\,$ and $\,\phi\,$ in terms of the boundary data are
\begin{equation}                                                  \label{eq:C-1-3-coeff-sol2ex}
T(x_1) \,=\, \overline{T} + \frac{2h}{k}\Big[\overline{q} +\frac{k'}{2h}  \Big(\overline{\phi}-\overline{\phi}_2 \Big)  \Big] \frac{e^{ax_1}-e^{ah}}{e^{ah}-e^{-ah}} \, ,                 \end{equation}
\begin{equation}                                                  \label{eqnarray:C-1-3-coeff-sol4ex}
\phi(x_1)\,=\, \overline{\phi} + \frac{2h}{k'}\Big[\overline{q} +\frac{k'}{2h}  \Big(\overline{\phi}-\overline{\phi}_2 \Big)  \Big]  \frac{e^{ax_1}-e^{ah}}{e^{ah}-e^{-ah}}  +  \frac{\overline{q}}{k'}(h-x_1)  \,    ,
\end{equation}
from which we deduce
\begin{equation}                                                  \label{eqnarray:C-1-3-coeff-sol2-h}
T(-h) \,=\,\overline{T} - 2hk^{-1}\overline{q} - k'k^{-1}(\overline{\phi}-\overline{\phi}_2)   \, .       
\end{equation} 
{\it Part $2$ of $BCP$ $II.1.3. \quad$} 
The remaining two equations, together with the appropriate boundary conditions, exactly coincide with $BCP$ $I.1.3$.
Hence, we can proceed as described at the end of Subsection \ref{subsubsection:SplittingProbB13}.

\begin{remark}
In order to avoid growth as $\;  h \rightarrow +\infty \;$ of the magnitude of the solution, we again assume (\ref{eq:assumption1}).
\end{remark}

\subsubsection{On controllability of temperature} 					              
By Eq. (\ref{eq:C-1-3-coeff-sol2ex}) we can deduce the following control property. 

\begin{remark} Let $\, -h \leq x_1 < h$.
For each choice of $\; \overline{u}_1,\,  \overline{u}_2,\,  \overline{u}_3, \, \overline{t}_1,\,  \overline{t}_2,\,  \overline{t}_3$, given any three quantities from $\, \left\{ \overline{T}, \,\overline{q} , \,  \overline{\phi} , \, \overline{\phi}_2 \right\}$, the remaining quantity can be choosen to control  $\,T(x_1)$. 
In particular, if  $\,\overline{T}\,$ and  $\,\overline{q}\,$ are assigned, then $\,T(x_1)\,$ is controllable by 
$\, \overline{\phi}_2-\overline{\phi}$.     
\end{remark}
\subsection{$BCP$ $I.3.3$, plate perpendicular to the polarization direction} 					              \label{subsubsection:ControlPr3.3}
Now consider a plate occupied by the same above material but having the polarization direction $\,x_3\,$ perpendicular to the plane of the plate. 
The plate is coated by an infinitesimally thin electrode on the plane $\,x_3=h$, so that all its mechanical effects may be ignored. 
Solutions of the form
\begin{equation}   \label{eq:soloftheform3x3}
 T=T(x_3) \,, \quad  \phi = \phi(x_3)\,, \quad 	u_i=u_i(x_3) \,,
\end{equation} 
satisfy (\ref{eqnarray:ConstEq1111})-(\ref{eq:bal355}) provided 
\begin{equation}             \label{eq:ConstEq1113+}    \nonumber
c_{44} u^{}_{1,\,33}                 
 \, =\,	0 \, , \qquad \qquad  c_{44} u^{}_{2,\,33} \, =\,	0 \, ,
\end{equation}
\begin{equation}             \label{eq:ConstEq1333333+}  \nonumber
c_{33} u^{}_{3,\,33}\,-\,\beta_{3}T_{,\,3} \,+\, e_{33}\phi_{, \,33} 
  \, =\,	0 \,, \qquad \qquad
\end{equation}
\begin{equation}             \label{eq:ConstEq11113+}
 e_{33}u^{}_{3,\,33} \,+\,
 \tilde{\omega}^{}_3  \,T_{,\,3} \, - \, \varepsilon_{33}\phi_{, \,33} \, =\,	0  \, ,
 \quad - \kappa^{}_{33} \, T_{,\,33} \,+\, \kappa^E_{33} \,\phi^{}_{,\,33}\,=\,0\,, 
\end{equation}
which are a system included in the general case considered in the Appendix.

$BCP$ $I.3.3:\;$
{\it To find the solution of the form (\ref{eq:soloftheform3x3}) to the field equations (\ref{eq:ConstEq1113+})-(\ref{eq:ConstEq11113+}),
subject to the ten boundary conditions	}
\begin{eqnarray}   \label{eqnarray:10bcI33} \nonumber
T(h) = \overline{T}  \, , \quad  \phi(h) = \overline{\phi} \,  ,    \quad   	
t_{3}(h) =\overline{t}_1 \,,  \quad t_{4}(h) =\overline{t}_2 \,, \quad t_{5}(h) =	\overline{t}_3 \,, \quad \\ 
 u^{}_{i}(-h)  = \overline{u}_i \quad(i=1,\,2,\,3) ,\quad 
 -D_{3}(-h) = \overline{D}   \,, \qquad 
 -q_{3}(-h) = \overline{q}  \,. \quad 
\end{eqnarray}
\subsubsection{General solution of $BCP$ $I.3.3$} 					              \label{subsubsection:gen-solution-P-IB+}

In particular, on setting in (\ref{eqnarray:systemgeneral}) 
\begin{equation}    \label{eq:eewith}
c=c_{33}\,, \; e=e'=e_{33} \, , \; \beta=\beta_3 \, , \; \omega=\tilde{\omega}^{}_3 \, , \; \varepsilon=\varepsilon_{33} \, , 
\; k=\kappa_{33} \, ,  \; k'=\kappa^E_{33} \, ,
\end{equation} 
we obtain Eq.s (\ref{eq:ConstEq1333333+})-(\ref{eq:ConstEq11113+}). Then
by Proposition \ref{proposition:pr1} the general solution to (\ref{eq:ConstEq1333333+})-(\ref{eq:ConstEq11113+})   is
\begin{equation}                                \label{eq:u12}
	u_1(x_3)=U_{11} x_3+U_{12} \,, \qquad u_2(x_3)=U_{21} x_3+U_{22}  \, ,
\end{equation}
\begin{equation}                                \label{eq:u32}
	u_3(x_3)=a^{-1}c^{-1}VT_1e^{ax_3}+U_{31} x_3+U_{32}   \, ,
\end{equation}
\begin{equation}                                \label{eq:T12}
	T(x_3)=T_1e^{ax_3}+T_2    \, ,
\end{equation}
\begin{equation}                                \label{eq:phi12}
	\phi(x_3)=KT_1e^{ax_3}+F_1 x_3+F_2    \, ,
\end{equation}
where 
$\;T_1, \, T_2, \,F_1, \, F_2, \, U_{\alpha 1}, \, U_{\alpha 2} \; \, (\alpha=1,\,2, \, 3)\;$
are arbitrary smooth functions of $\,\tau\,$  and the notation (\ref{eq:eewith}) and (\ref{equation:a}) is used.  Moreover, the constant $\,a$, given by 
\begin{equation}              \label{eq:assumption2}
\qquad \qquad \qquad \qquad \qquad  a:= \frac{\beta_{3} e_{33} + c_{33} \tilde\omega_{3}}{K(e_{33}^2+c_{33} \varepsilon_{33})} \,  ,
\end{equation}
 is supposed positive.
\subsubsection{Decomposition of $BCP$ $I.3.3$} 					              \label{subsubsection:SplittingProbB13+}
 We solve $BCP$ $I.3.3$ by decomposing it into  two parts.
{\it Part $1$ of $BCP$ $I.3.3. \quad$} 
First, we note that Eq.s (\ref{eqnarray:DDD})$_3$,  (\ref{eq:u32})--(\ref{eq:phi12})  and (\ref{equation:a}) imply
\begin{equation}                                                 \label{eq:d333}
D_3(x_3)\,=\, \omega T_1 \, e^{ax_3}+\omega T_2 + e_{}U_{31}- \varepsilon_{}F_1  \,,
\end{equation}
and consequently, the solutions (\ref{eq:u32})--(\ref{eq:phi12}) meet the boundary conditions (\ref{eqnarray:6bc}) when
 \begin{equation}  \label{eq:ctauugsBàò}
T_1\, e^{ah} \,+\, T_2 \, =\,  \overline{T} \,   ,  \qquad 
      \,K \,T_1\, e^{ah } \,+\,F_1 \, h \,+\,F_2 \,=\,\overline{\phi} \, ,  
\end{equation}
\begin{equation}                                                  \label{eq:bc3333implysu3Bàò}
-\beta T_2 + eF_1+cU_{31} \, =\, \overline{t}_1  \, , \qquad
a^{-1}c^{-1}V T_1 \, e^{-ah}-U_{31} h + U_{32} \,=\, \overline{u}_3 \, ,
\end{equation}
\begin{equation}                                                 \label{eq:ccoollxxbisBàò}
-\omega T_2 -e_{}U_{31}+ \varepsilon_{}F_1 =\overline{D}  \,, \qquad
-\kappa_{}K^{-1}F_1=\overline{q}          \, .
\end{equation}
By solving the above system of equations in the unknowns 
$\,(T_1, \, T_2 , \, F_1, \, F_2 , \, U_{3 \, 1}, \, U_{3 \, 2})$,
we obtain
\begin{equation}          \label{equation:coeffPIB+1}              
 T_2 \,=\,-\frac{ 
 e\overline{t}_1\,+\,c\overline{D}\,+\,Kk^{-1} (e^2+c \varepsilon) \, \overline{q}}{(\beta e + c \omega)e^{ah}} \,  ,
 \qquad  T_1 \,=\, e^{-ah} \, ( \overline{T} \,-\, \, T_2) \, , 
\end{equation} 
 \begin{equation}  \label{equation:coeffFF+} 
  F_1 \,=\,- \frac{\overline{q}}{k'} \, , \qquad 
 F_2 \,=\, \overline{\phi} - h F_1 - K e^{ah} T_1    \, ,      \end{equation} 
 \begin{equation}  
U_{31}=\frac{\omega \overline{t}_1-\beta \overline{D} + k'^{-1}(e \omega - \beta \varepsilon) \overline{q}}{\beta e + c\omega}  , \;
 U_{32}=\overline{u}_3 + h U_{31}  - \frac{V}{ace^{ah}} T_1     \, , 
\end{equation}
which on substitution in the general solution (\ref{eq:T12})-(\ref{eq:phi12}) yields
the solution to the first part of  $BCP$ $I.1.3$.  In particular, we have 
\begin{equation}               	   \label{eq:TTFFB+}           
\phi(x_3)\,=\,K T_1 \,\Big( e^{ax_3}- e^{ah}\Big)\,-\, k'^{-1}\, \overline{q} (x_3-h)\,+\, \overline{\phi} \, , 
\end{equation}
\begin{equation}               	   \label{eq:TTFFBt+}         
T(x_3) \,=\, T_2 \,\Big( 1-e^{a(x_3-h)}\Big) \,+\, \overline{T} \,e^{a(x_3-h)}\, ,
\end{equation}
and thus, by  (\ref{equation:coeffPIB+1}) too, we have
\begin{equation}               	   \label{equation:Phi-h+}           
\phi(-h)=K \Big[\overline{T}\,+\,\frac{ 
 e\overline{t}_1\,+\,c\overline{D}\,+\,k'^{-1} (e^2+c \varepsilon) \, \overline{q}}{(\beta e + c \omega)e^{ah}} \,\Big]\Big(e^{-2ah}-1\Big) +\frac{2h}{k'}\overline{q}+ \overline{\phi} \, , 
\end{equation}
\begin{equation}               	   \label{equation:TTFFBct-h+}           
T(-h) \,=\,\frac{ 
 e\overline{t}_1\,+\,c\overline{D}\,+\,k'^{-1} (e^2+c \varepsilon) \, \overline{q}}{(\beta e + c \omega)e^{ah}} \,
\Big( e^{-2ah}- 1\Big)\, + \, e^{-2ah}\, \overline{T} \, . 
\end{equation}
For $\; 0=\overline{t}_1=\overline{D}=\overline{q}\;$  we thus obtain 
\begin{equation}               	              \label{equation:Phi-h00+}           
\phi(-h)= K \Big(e^{-2ah}-1\Big)\, \overline{T} + \overline{\phi} \, , 
\qquad T(-h) \,=\, \, e^{-2ah}\, \overline{T} \, , 
\end{equation}
which yields  electric potential and temperature in the plane $\,x_1=-h\,$ in terms of  electric potential and temperature $\, \overline{\phi}\,$, $\,\overline{T}\,$ at $\,x_1=h$.

{\it Part $2$ of $BCP$ $I.3.3. \quad$} 
The remaining two equations (\ref{eq:ConstEq1113+}), subject to the remaining four boundary conditions (\ref{eq:bb1bx}), can be solved exactly as before in Subsection \ref{subsubsection:SplittingProbB13}.

 \begin{remark}        \label{remark:133}
We point out that in order to avoid growth as $\;  h \rightarrow +\infty \;$ of the magnitude of the gradient
$$ |  \nabla {\bf u}| + |\nabla T |+ |\nabla \phi | \, ,$$
of any solution  $\; ({\bf u}, \; T , \; \phi ) \; $ 
to $BCP$ $I.1.3$, we assume 
\begin{equation}              \label{eq:assumption2b}
\qquad \qquad \qquad \qquad \qquad  a:= \frac{\beta_{3} e_{33} + c_{33} \tilde\omega_{3}}{K(e_{33}^2+c_{33} \varepsilon_{33})} > 0 \; .	
\end{equation}
\end{remark} 

\subsection{On controllability in $BCP$ $I.3.3$} 					              \label{subsubsection:Control-I.3.3}
By Eq.s (\ref{eq:TTFFB+}), (\ref{eq:TTFFBt+}) we can deduce the following control property.
 \begin{remark} 
Let $\, -h \leq x_3 < h$.
For each choice of $\, \overline{u}_1, \, \overline{u}_2, \, \overline{u}_3, \, \overline{t}_2,\, \overline{t}_3\,$ and $\, \overline{\phi}$, given any three quantities from $\, \left\{\overline{t}_1 , \, \overline{D} , \, \overline{q}, \, \overline{T} \right\} \,$, the fourth quantity can be choosen to control  either $\,T(x_3)\,$ or $\,\phi(x_3)$. 
\end{remark}   

\subsection{$BCP$ $II.3.3$, plate perpendicular to the polarization direction} 					              \label{subsubsection:PrC3.3}
Here $\,{\mathcal P}\,$ (see $BCP$ $I.3.3$) is bounded by the parallel planes $\,x_3= \pm h\,$ on which are coated two infinitesimally thin electrodes whose mechanical effects therefore may be ignored. We seek solutions of the form (\ref{eq:soloftheform3x3})
which after substitution in  (\ref{eqnarray:ConstEq1111})-(\ref{eq:bal355}) give Eq.s  (\ref{eq:ConstEq1113+})-(\ref{eq:ConstEq11113+}).

$BCP$ $II.3.3\,$:$\;$
{\it To find the solution of the form (\ref{eq:soloftheform3x3}) to the field equations (\ref{eq:ConstEq1113+})-(\ref{eq:ConstEq11113+}),
which satisfies the ten boundary conditions}
\begin{eqnarray}   \label{eqnarray:10bcII33}  \nonumber
T(h) = \overline{T}  \, , \quad  \phi(h) = \overline{\phi} \,  ,    \quad   	
t_{3}(h) =\overline{t}_1 \,,  \quad t_{4}(h) =\overline{t}_2 \,, \quad t_{5}(h) =	\overline{t}_3 \,, \quad \\ 
u^{}_{i}(-h)  = \overline{u}_i \quad(i=1,\,2,\,3) ,\quad 
\phi(-h) = \overline{\phi}_2   \,, \qquad 
 -q_{3}(-h) = \overline{q}  \,. \quad\quad
\end{eqnarray}
\subsubsection{General solution of $BCP$ $II.3.3$} 					              \label{subsubsection:gen-solution-P-IB+c}

The equations correspponding to (\ref{eq:ConstEq1113+})-(\ref{eq:ConstEq11113+}) are of the form  (\ref{eqnarray:systemgeneral}) on setting 
\begin{equation}    \label{eq:eewithgg}
c=c_{33}\,, \; e=e'=e_{33} \, , \; \beta=\beta_3 \, , \; \omega=\tilde{\omega}^{}_3 \, , \; \varepsilon=\varepsilon_{33} \, , 
\; k=\kappa_{33} \, ,  \; k'=\kappa^E_{33} \, ,
\end{equation} 
so that by Proposition \ref{proposition:pr1} the general solution to Eq.s (\ref{eq:ConstEq1113+})-(\ref{eq:ConstEq11113+})  is given by
Eq.s (\ref{eq:u12})-(\ref{eq:phi12}),
where 
$$T_1, \, T_2, \,F_1, \, F_2, \, U_{\alpha 1}, \, U_{\alpha 2} \; \, (\alpha=1,\,2, \, 3)$$
are arbitrary smooth functions of $\,\tau\,$  and we adopt the notation defined in Eq.s (\ref{eq:eewith}), (\ref{equation:a}).
\subsubsection{Decomposition of $BCP$ $II.3.3$} 					              \label{subsubsection:SplittingProbB13+pp}
 We solve $BCP$ $II.3.3$ as follows:

{\it Part $1$ of $BCP$ $II.3.3. \quad$} 
The general solution (\ref{eq:u32})--(\ref{eq:phi12}) to Eq.s  (\ref{eq:ConstEq1333333+})--(\ref{eq:ConstEq11113+}) satisfies the six boundary conditions
\begin{eqnarray}    \label{eqnarray:6bcII.3.3} \nonumber
	T(h)=\overline{T} \,, \quad  \phi(h)=\overline{\phi}  \,, \quad t_3(h)=\overline{t}_1 \,, \\ 
u_3(-h)=\overline{u}_3 \, , \quad  \phi(-h)=\overline{\phi}_2 \,, \quad -q_{1}(-h)=\overline{q} \, 
\end{eqnarray}
provided that
 \begin{equation}  \label{eq:ctauugsBààà}
T_1\, e^{ah} \,+\, T_2 \, =\,  \overline{T} \, ,      \qquad 
      \,K \,T_1\, e^{ah } \,+\,F_1 \, h \,+\,F_2 \,=\,\overline{\phi} \, ,
\end{equation}
\begin{equation}                                                  \label{eq:bc3333implysu3Bàkmn}
- \beta T_2 + eF_1+cU_{31} \, =\, \overline{t}_1  \,, \qquad 
a^{-1}c^{-1}V T_1 \, e^{-ah}-U_{31} h + U_{32} \,=\, \overline{u}_3 \, ,
\end{equation}
\begin{equation}                                                 \label{eq:ccoollxxbisBàkjh}
\,K \,T_1\, e^{-ah } \,-\,F_1 \, h \,+\,F_2 \,=\,\overline{\phi}_2 \, , \qquad
-\kappa_{}K^{-1}F_1=\overline{q}          \,,
\end{equation}
which can be solved for the unknowns 
$$(T_1, \, T_2 , \, F_1, \, F_2 , \, U_{3 \, 1}, \, U_{3 \, 2}  ) $$
to give in particular the expressions 
\begin{equation}                                            \label{eq:C-1-3-coeff-solT1n}
T_1=\frac{2h}{k(e^{ah}-e^{-ah})}\Big[\overline{q} +\frac{k'}{2h}  \Big(\overline{\phi}-\overline{\phi}_2 \Big)  \Big]   \, ,
   \end{equation}
and (\ref{eq:C-1-3-coeff-solT1T2})$_{2}$, (\ref{eq:C-1-3-coeff-sol3}).
Hence, (\ref{eq:T12}) and (\ref{eq:phi12}) become
\begin{equation}                                                  \label{eq:C-1-3-coeff-sol23}
T(x_3) \,=\, \overline{T} + \frac{2h}{k}\Big[\overline{q} +\frac{k'}{2h}  \Big(\overline{\phi}-\overline{\phi}_2 \Big)  \Big] \frac{e^{ax_3}-e^{ah}}{e^{ah}-e^{-ah}}          \, ,        \end{equation}
\begin{equation}                                                  \label{eqnarray:C-1-3-coeff-sol43}
\phi(x_3)\,=\,\frac{2h}{k'}\Big[\overline{q} +\frac{k'}{2h}  \Big(\overline{\phi}-\overline{\phi}_2 \Big)  \Big]\frac{e^{ax_3}-e^{ah}}{e^{ah}-e^{-ah}} + \overline{\phi} + \frac{\overline{q}}{k'}(h-x_3)  \,    ,
\end{equation}    
and we conclude that
\begin{equation}                                                  \label{eqnarray:C-1-3-coeff-sol2-h3}
T(-h) \,=\,\overline{T} - 2hk^{-1}\overline{q} - k'k^{-1}(\overline{\phi}-\overline{\phi}_2)   \, .       
     \end{equation}

{\it Part $2$ of $BCP$ $II.3.3. \quad$} 
The remaining two equations, subject to the appropriate boundary conditions, exactly coincide with the corresponding equations of $BCP$ $I.3.3$.
Hence we can solve them by the method of Subsection \ref{subsubsection:SplittingProbB13}.

\begin{remark}
In order to avoid growth as $\;  h \rightarrow +\infty \;$ in the magnitude of the gradient of the solution, we assume (\ref{eq:assumption2b}).
\end{remark}

\subsection{On controllability of temperature}     \label{subsubsection:Control-II.3.3}
By Eq. (\ref{eq:C-1-3-coeff-sol23}) we can deduce the following control property.
 \begin{remark} 
Let $\, -h \leq x_3 < h$.
For each choice of $\, \overline{\bf t}\,$ and $\, \overline {\bf u}$, given any three quantities from $\, \left\{ \overline{T}, \,\overline{q} , \,  \overline{\phi} , \, \overline{\phi}_2 \right\}$, the remaining quantity can be choosen to control  $\,T(x_3)$. 
    In particular, when $\,\overline{T}, \,\overline{q}\,$ are assigned,  $\,T(x_3)\,$ is controllable by
$\, \overline{\phi}-\overline{\phi}_2$.     
\end{remark}
\section{Conclusions and perspectives} 					              \label{section:concl}
We have shown that, for a piezothermoelastic plate referred to a natural configuration, in the presence of a quasi-static incremental temperature given on one of its bounding faces, on the other bounding face either the electric potential or the temperature can be controlled by certain boundary data.

An aim of a future investigation could be to examine how these result generalize when the initial configuration of the plate is not a natural configuration, that is, when there is some initial mechanical, thermal and/or electric field.  

\section{Appendix}  					   \label{section:ed1lin}
The following elementary result on first order differential equations is used.

\begin{remark} \label{remark:uno} If $\, f=f(x) \,$ is a scalar function of the real variable $\,x\,$ and $\,a,\,b \in I\!\!R$, then   
the general solution of the linear first-order differential equation
$$ f'=a(f-b)$$
is $\;f = \gamma e^{ax}+b$,          
where $\,\gamma\,$ is an arbitrary real constant.
\end{remark}
\begin{proposition}  					   \label{proposition:pr1}
Let  $\,c,\,e,\,e', \, \beta, \, \omega, \, \varepsilon, \, k, \, k' \,$ be real scalars.  Then the system of linear differential equations  

\begin{eqnarray}               	   \label{eqnarray:systemgeneral}                  \nonumber
	cu_{,\,xx}-\beta T_{,\,x} +e'\phi_{,\,xx}=0                               \\
	eu_{,\,xx}+\omega T_{,\,x}-\varepsilon \phi_{,\,xx}=0       \\
	-kT_{,\,xx}+k' \phi_{,\,xx}=0                                           \nonumber
\end{eqnarray}
in the unknown scalar functions $$\; T=T(x), \quad  \phi=\phi(x), \quad  u=u(x) \, ,  $$ of the real variable $\,x$, 
has the general solution
\begin{eqnarray}               	   \label{eqnarray:SystemGenSol}            \nonumber
T(x)=T_1 \, e^{ax}+T_2 \,                                           \\
\phi(x)=K T_1 \, e^{ax}+F_1 x +F_2  \,    \\
	u=a^{-1}c^{-1}VT_1   e^{ax}+U_1x+U_2 \,                              \nonumber
\end{eqnarray}
where $$\left( \,T_1, \, T_2, \,F_1, \, F_2, \,U_1, \, U_2  \right) \in I\!\!R^6$$
are arbitrary and 
\begin{equation}    \label{equation:a}
K:=k/k'  , \; A:= \beta e + c\omega  , \;  B:= ee'+c \varepsilon , \; 	a:=AK^{-1}B^{-1}
, \; V:=\beta -Kae'\, . 
\end{equation}
\end{proposition}  		
{\bf Proof.} $\quad$ 
Equation (\ref{eqnarray:systemgeneral})$_3$  yields $\;\phi_{,\,xx}=KT_{,\,xx}$, 
thus
\begin{equation}  \label{equation:ff}
\phi_{,\,x}=K T_{,\,x} + F_1 \,,  \quad F_1 \in I\!\!R \, ,
\end{equation}

and Eq.s (\ref{eqnarray:systemgeneral})$_{1,\,2}$ become
\begin{eqnarray}               	   \label{eqnarray:sg2}                  \nonumber
	u_{,\,xx}=c^{-1}\big( \beta T_{,\,x}-e'KT_{,\,xx}\big)  \, , \qquad
	u_{,\,xx}=e^{-1}\big( -\omega T_{,\,x}+\varepsilon KT_{,\,xx}\big)    \,.    
\end{eqnarray}
By eliminating  $\,	u_{,\,xx}\,$ from these two equalities, we obtain the second-order equation 
\begin{equation}    \label{equation:TTT}
T_{,\,xx}=aT_{,\,x} \,,
\end{equation}
with $\,a\,$ defined in (\ref{equation:a}); consequently
$\; T_{,\,x}=a(T-T_2) \; $
 where $\,T_2 \,$ is an arbitrary constant.
 By Remark \ref{remark:uno} the latter equation has the general solution
\begin{equation}   \label{equation:solgenT}
T(x)=T_1 e^{ax}+T_2 \,  \qquad  	(T_1 , \, T_2) \in I\!\!R^2 \, ,
\end{equation}
which by substitution in  (\ref{equation:ff}) enables us to conclude that (\ref{eqnarray:SystemGenSol})$_{2}$ holds. 
Lastly, insertion of the expressions for $\,T\,$ and $\, \phi\,$ into (\ref{eqnarray:sg2})$_1$ yields	$\;u_{,\,xx}=c^{-1} T_1 V a e^{ax} \;$
where $\; V:=\beta -Ke'a$.
Hence by integration we obtain $\,	u_{,\,x}=c^{-1} T_1 V e^{ax}+U_1 $, which yields (\ref{eqnarray:SystemGenSol})$_3$. $\quad \diamondsuit$
\section{Acknowledgments}
The author would like to thank Professor R. Knops for his discussion and suggestions on the present paper.

\end{document}